%% file: main.tex
\documentclass[twocolumn]{aastex62}

\usepackage[utf8]{inputenc}
\DeclareUnicodeCharacter{2212}{-}
\usepackage{graphicx}
\graphicspath{{./}}
\usepackage{amsmath}
\usepackage{afterpage}
\usepackage{enumitem}
\usepackage{xcolor}
\usepackage{float}
\setenumerate{itemsep=0mm}

\newcommand{\stamp}[0]{Eri~IV\xspace}

\input{commands}

\shorttitle{Discovery of an Ultra-Faint Dwarf Galaxy Candidate in Eridanus}

\shortauthors{Cerny et al.}

\begin{document}

\reportnum{\footnotesize 
FERMILAB-PUB-21-319-E}

\title{Eridanus IV: an Ultra-Faint Dwarf Galaxy Candidate Discovered in the DECam Local Volume Exploration Survey}

\input{authors}

\begin{abstract}
We present the discovery of a candidate ultra-faint Milky Way satellite, Eridanus IV (DELVE J0505$-$0931), detected in photometric data from the DECam Local Volume Exploration survey (DELVE). Eridanus IV is a faint ($M_V = -4.7 \pm 0.2$), extended ($r_{1/2} = 75^{+16}_{-13} \pc$), and elliptical ($\epsilon = 0.54 \pm 0.1$) system at a heliocentric distance of $76.7^{+4.0}_{-6.1} \kpc$, with a stellar population that is well-described by an old, metal-poor isochrone (age of $\tau \sim 13.0  \Gyr$ and metallicity of ${\rm [Fe/H] \lesssim -2.1}$ dex). These properties are consistent with the known population of ultra-faint Milky Way satellite galaxies. Eridanus IV is also prominently detected using proper motion measurements from \Gaia Early Data Release 3, with a systemic proper motion of $(\mu_{\alpha} \cos \delta, \mu_{\delta}) = (+0.25 \pm 0.06, -0.10 \pm 0.05)$ mas yr$^{-1}$ measured from its horizontal branch and red giant branch member stars. We  find that the spatial distribution of likely member stars hints at the possibility that the system is undergoing tidal disruption. 

\end{abstract}



\section{Introduction}
\label{sec:intro}

\par The \LCDM cosmological model successfully predicts structure formation across a wide range of mass scales, from the most massive galaxy clusters to the lowest mass dwarf galaxies. At the low-mass, low-luminosity extreme (stellar masses $M_{*} \lesssim 10^5 \Msun$, $M_V \gtrsim -7.7$; \citealt{Simon:2019}) are the so-called     ``ultra-faint'' dwarf galaxies, which have only been detected in the very local Universe as satellites of the Milky Way and other nearby galaxies. In addition to being ideal systems for testing the hierarchical model of structure formation at the smallest scales, these systems are also pristine laboratories for studying the nature of dark matter \citep[e.g.,][]{Nadler:2020b}, the synthesis of heavy elements \citep[e.g.,][]{Ji:2016}, and the physics of reionization \citep[e.g.,][]{2017MNRAS.469L..83W}. 
\par Despite the scientific insights to be gained from studying these faint systems, the known population of Milky Way satellite galaxies was limited to roughly a dozen brighter, ``classical'' dwarf spheroidal galaxies throughout much of the last century. 
However, in the last two decades, successive efforts by large, multi-band digital sky surveys (e.g., SDSS, PanSTARRS, DES, and HSC-SSP) have extended observational sensitivity to much fainter systems, expanding the census of known Milky Way satellite galaxies to $\roughly 60$ systems \citep[][and references therein]{Drlica-Wagner:2020}. Moreover, the \Gaia satellite has provided precise proper motion measurements of distant stars, enabling detailed kinematic and orbital measurements of satellite galaxies \citep[e.g.,][]{Fritz2018A&A...619A.103F, Erkal2020MNRAS.495.2554E} and the discovery of new ultra-faint systems with algorithms that leverage proper motion information to identify gravitationally-bound stellar systems \citep[e.g.,][]{Torrealba:2019,Darragh-Ford:2020}.

\begin{figure*}
\includegraphics[width=\textwidth]{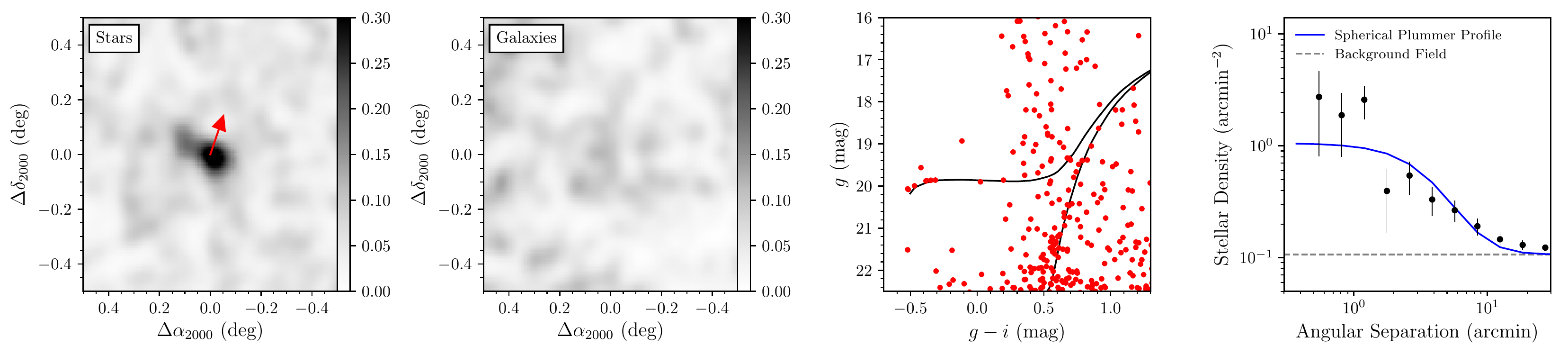}
\caption{Diagnostic plots similar to those that were visually inspected for candidates identified by the \code{simple} search. (Left) Isochrone-filtered stellar density field around Eri~IV, smoothed with a $2\arcmin$ Gaussian kernel. Eri IV is visible as a prominent overdensity against the background stellar field. The color scale  has been set to highlight the candidate tidal feature to the northeast of Eri~IV (\secref{discussion}). The direction of the solar-reflex-corrected proper motion vector of the system is shown as a red arrow (\secref{proper_motion}). (Center Left) Isochrone--filtered galaxy density field, again smoothed with a $2\arcmin$ Gaussian kernel. (Center Right) Color--magnitude diagram of stars selected within $\sim 2$ half-light radii of Eri~IV. A significant number of stars lie close to the best-fit isochrone (\secref{results}), including five likely blue horizontal-branch stars and a number of lower red giant branch stars. (Right) Radial stellar density profile centered at the location of Eri~IV, assuming spherical symmetry. The blue curve represents a Plummer profile with scale radius of $a_{\rm h} = 4.9'$ (\secref{results}). The apparent discontinuity in the density profile at $\sim 1$ arcmin is due to the slight miscentroiding introduced by the candidate tidal feature and random statistical variations due to the small number of stars.}
\label{fig:diagnostic}
\end{figure*}

\par In this \textit{Letter}, we report the discovery of a new ultra-faint stellar system in the constellation Eridanus, identified through a matched-filter search of data from the DECam Local Volume Exploration survey \citep[DELVE;][]{Drlica-Wagner:2021}. The system's size and luminosity are consistent with the known population of ultra-faint dwarf galaxies orbiting the Milky Way. Based on the high likelihood that this system is a dwarf galaxy, rather than a star cluster, we adopt the dwarf galaxy naming convention and refer to it as Eridanus IV (Eri~IV) after the constellation within which it is found.

\section{DELVE Data and Satellite Search}
\subsection{DELVE Data}
\label{sec:data}

DELVE is an ongoing observational campaign using the Dark Energy Camera  \citep[DECam;][]{Flaugher:2015} on the 4-meter Blanco telescope to study the satellite systems of the Milky Way, Magellanic Clouds, and four Magellanic-analogs within $\roughly 2 \Mpc$ \citep{Drlica-Wagner:2021}. DELVE seeks to complete observational coverage of the high-Galactic-latitude ($|b| > 10\degree$) southern sky in the $g,r,i,z$ bands by combining 126 nights of observing with all existing public DECam community data with exposure times $>30$ seconds. More detailed information about the observational strategy and scientific goals of DELVE can be found in \citet{Drlica-Wagner:2021}.
\par We used an updated internal data release from DELVE constructed from $\roughly 20{,}000$ $g$- and $i$-band exposures covering $\roughly 6{,}400 \deg^{2}$ in the southern Galactic cap outside the DES footprint. All exposures were processed with the DESDM pipeline \citep{Morganson:2018}, which utilizes \code{SourceExtractor} and \code{PSFEx} \citep{Bertin:1996,2011ASPC..442..435B} for automatic source detection and photometric measurement. Astrometry was calibrated against \Gaia DR2, which provides $\roughly 30$ mas relative astrometric calibration. The DELVE photometry is calibrated for each DECam CCD by matching stars to the ATLAS Refcat2 catalog \citep{2018ApJ...867..105T}. A multi-band catalog of unique astronomical objects was generated by matching the \code{SourceExtractor} catalogs from each CCD image as described in \citet{Drlica-Wagner:2021}. 
For each source, we calculated interstellar extinction due to foreground dust by interpolating the $E(B-V)$ maps of \citet{Schlegel:1998} with the rescaled normalization from \citet{2011ApJ...737..103S} and the reddening coefficients from Section 4.2 of \citet{DES:2018}. Hereafter, all quoted magnitudes are corrected for extinction. 
\par We separate stars and galaxies using the \code{Source{\allowbreak}Extractor} \var{SPREAD\_MODEL} parameter.
When performing our search, we selected a relatively pure sample of stars, $|\var{SPREAD\_MODEL\_G}| < 0.003 + |\var{SPREADERR\_MODEL\_G}|$, and restricted our stellar sample to $g < 23.0$ mag, which corresponds to the median $g$-band magnitude limit of the DELVE data at S/N=10.
To characterize Eri~IV, we selected a more complete sample of stars following Section 4.8 of \citet{Drlica-Wagner:2021}, $0 \leq \var{EXTENDED\_CLASS\_G} \leq 2$, which has been found to be $>95\%$ complete down to a magnitude of $g \sim 22.5 \magn$.
We use this value as the $g$-band magnitude limit for our characterization analysis, paired with the $i$-band S/N=10 magnitude limit of $i\sim22.7 \magn$. 
We find that the derived properties of Eri~IV are insensitive to moderate ($\pm 0.3$ mag) variations in the assumed magnitude limits.

\subsection{Matched-Filter Search}
\begin{figure*}
\center
\includegraphics[width=\textwidth]{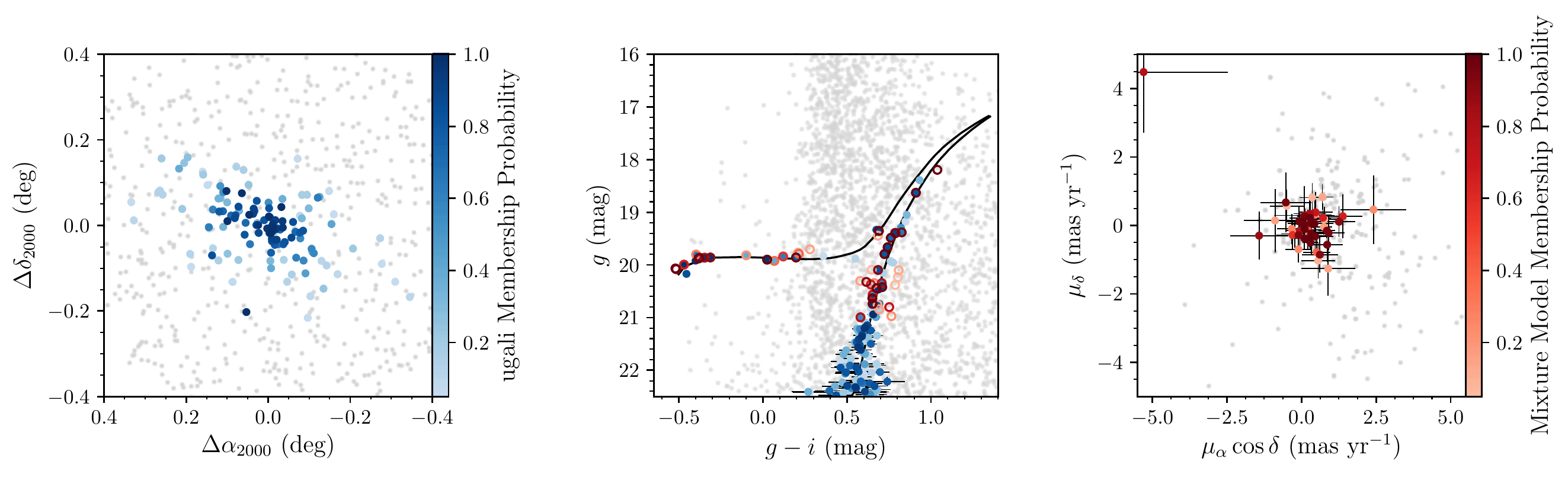}
\caption{(Left) Spatial distribution of isochrone-filtered stars in a small region around Eri~IV. Stars with \code{ugali} membership probability greater than 5\% ($p_{\code{ugali}} > 0.05$) are colored by their respective probabilities, while those below this threshold are colored gray. (Center) Color--magnitude diagram of all sources within a $30\arcmin$ radius ($\sim 9\, r_{\rm h}$). Member stars are colored in the same way as the lefthand panel, and stars identified in the \Gaia data are outlined with circles, where the colors of the circles represent the mixture model probability ($p_{\rm MM}$) of each star (see right panel). (Right) Proper motion scatterplot of all sources passing the cuts described in \secref{proper_motion}. Sources with $p_{\rm MM} < 0.05$ are colored in gray, while those with $p_{\rm MM} > 0.05$ are colored by their mixture model membership probabilities.}
\label{fig:membership}
\end{figure*}

We searched for old, metal-poor stellar systems in the DELVE catalog using the \code{simple}\footnote{\url{https://github.com/DarkEnergySurvey/simple}} matched-filter algorithm, which has been applied to data from several DECam survey programs to help discover more than 20 Milky Way satellites \citep{Bechtol:2015, Drlica-Wagner:2015,Mau:2019,MauCerny2020,delve2}. In brief, \code{simple} uses an isochrone matched-filter approach in color--magnitude space to increase the contrast of halo substructure relative to foreground Milky Way stars at a given distance. Specifically, we split the DELVE catalog into \code{HEALPix} pixels \citep{Gorski:2005} at \code{nside} = 32 ($\sim 3.4$ deg$^{2})$ and searched for stars consistent with a PARSEC \citep{Bressan:2012} isochrone of age $\tau = 12$ Gyr and metallicity $Z = 0.0001$ ($\rm [Fe/H] \sim -2.2$ dex). We scanned this isochrone over a range of distance moduli from $16.0 \leq m-M \leq 23.5$ mag in steps of 0.5 mag. For each step in distance, we selected stars consistent with the isochrone according to $\Delta (g-i) < \sqrt{0.1^{2} + \sigma_{g}^{2} + \sigma_{i}^{2}}$, and then smoothed the resulting density field with a $2'$ Gaussian kernel. Local spatial overdensities were identified by iteratively raising a density threshold until fewer than ten disconnected peaks remained \citep{Bechtol:2015}. The resulting candidates with significance $>5\sigma$ were visually inspected.
\par Eri~IV was the most promising new candidate identified by the \code{simple} search.\footnote{Ongoing searches of the DELVE catalog (including searches leveraging $r$-band data) may uncover additional stellar systems. The presentation of just one system should not be interpreted as a statement of detection completeness for the DELVE data.} The corresponding density peak had a Poisson significance of 11.8$\sigma$ relative to the stellar density in an annular background region, and showed evidence of a prominent blue horizontal branch, lending significant credence to the detection.
We note that this candidate system resides within the Pan-STARRS1 DR1 (PS1) footprint, but it was not identified in the census conducted by \citet{Drlica-Wagner:2020}. However, the observational selection function constructed by that work suggests that the detectability of a system with the properties of Eri~IV is only $\roughly 25\%$ using the PS1 data, and thus the previous non-detection is not unexpected (see \secref{results}).

\par In \figref{diagnostic}, we present diagnostic plots similar to those inspected during the search process. These include the smoothed stellar and galaxy density around the candidate, along with a color--magnitude diagram, and a (azimuthally-averaged) radial profile centered at the location of the candidate system.

\section{Properties of Eridanus IV}
\label{sec:results}

\subsection{Morphology and Stellar Population}

\par We fit the morphological and stellar population parameters of \stamp using the unbinned maximum-likelihood formalism implemented in  
the Ultra-faint Galaxy Likelihood (\ugali) software toolkit \citep{Bechtol:2015,Drlica-Wagner:2020}.\footnote{\url{https://github.com/DarkEnergySurvey/ugali}} 
We modeled the 2D spatial distribution of stars with a Plummer profile, and fit the system's color--magnitude diagram with a set of PARSEC \citep{Bressan:2012} isochrones assuming a \citet{Chabrier:2001} initial mass function. We  simultaneously constrained the centroid (\ra, \dec), semi-major axis ($a_{\rm h}$), ellipticity ($\epsilon$), and position angle (P.A.) of the Plummer profile, and the age ($\tau$), metallicity ($Z$), and distance modulus ($m-M$) of the isochrone by sampling the posterior probability distributions of each parameter using an affine-invariant Markov Chain Monte Carlo (MCMC) ensemble sampler \citep[\code{emcee};][]{Foreman-Mackey:2013}. We then derived estimates for the system's azimuthally averaged angular and physical half-light radii ($r_{\rm h}$ and $r_{1/2}$, respectively), absolute visual magnitude ($M_V$), average surface brightness within the half-light radius ($\mu$), mean metallicity (${\rm [Fe/H]}$), and the total stellar mass of the system ($M_{*}$). 
\par We calculated the \code{ugali} membership probability for each star based on the Poisson probability that it belongs to Eri~IV based upon its spatial position, flux, and photometric uncertainty, given an empirical model of the local stellar field population and of a putative dwarf galaxy.  We also report the quantity $\sum p_{i,\code{ugali}}$, 
which can be interpreted as the number of Eri IV member stars brighter than the DELVE magnitude limits.
\par The resulting values for each parameter are shown in \tabref{properties}.  We note that the uncertainties for the system's age and distance modulus are significantly asymmetric due to a bimodal posterior distribution. We believe that this behavior is likely introduced by foreground stars just brighter than the horizontal branch and do not attribute it to the presence of a second stellar population. We present the MCMC-derived posterior probability distributions for all parameters in \appref{posterior}.

\input{Table1}

\begin{figure*}
    \centering
    \includegraphics[width = .95\textwidth]{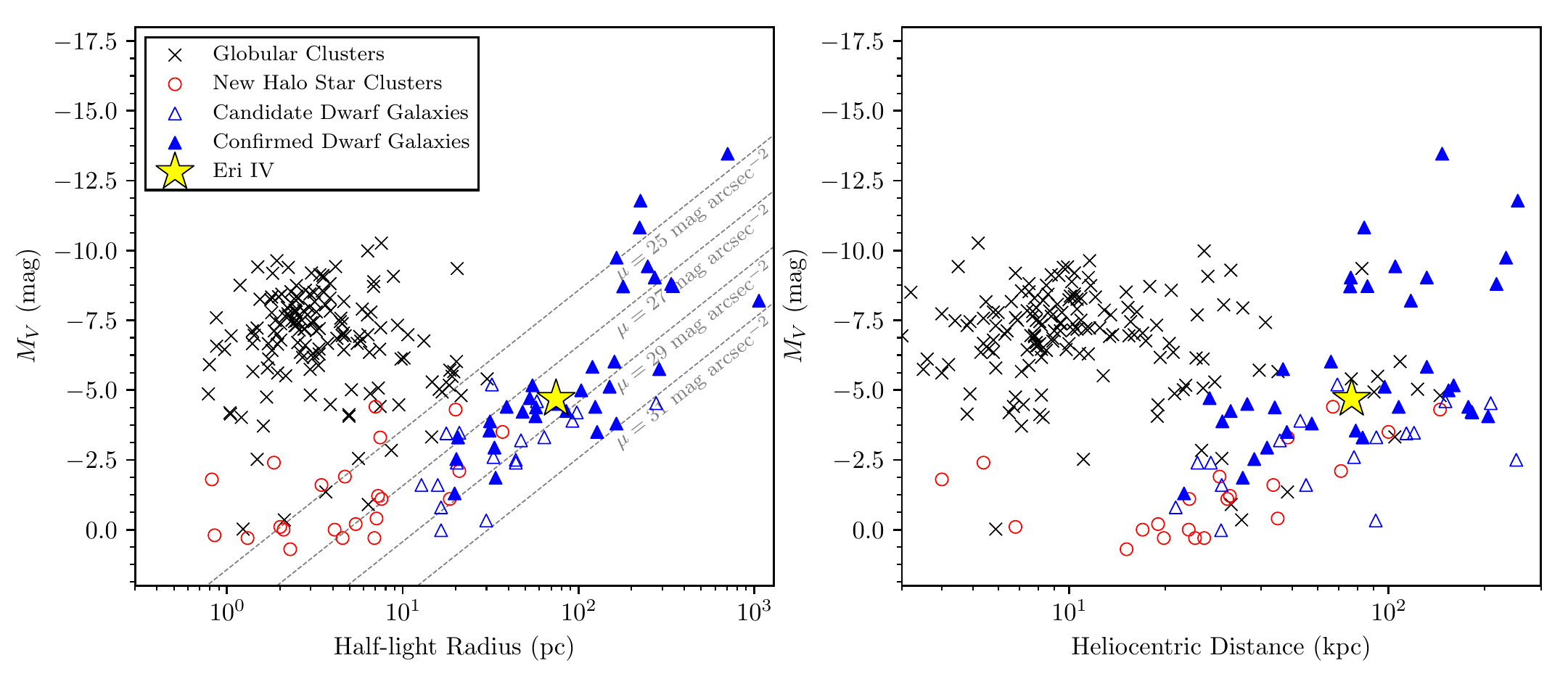}
    \caption{(Left) Absolute $V$-band magnitude ($M_V$) vs.\ azimuthally-averaged physical half-light radius ($r_{1/2}$) for the population of known Milky Way satellites, including classical globular clusters (black crosses; \citealt[][2010 edition]{Harris:1996}), faint halo star clusters (red circles; \citealt[][and references therein]{delve2}), and confirmed and candidate dwarf galaxies (filled and unfilled blue triangles, respectively; \citealt[][and references therein]{Drlica-Wagner:2020}). Eri~IV (yellow star) is consistent with the population of known dwarf galaxies. The dashed gray lines indicate constant surface brightness, as labeled.  (Right) Absolute $V$-band magnitude vs.\ heliocentric distance for the same systems.}
    \label{fig:population}
\end{figure*}

\subsection{Proper Motion}
\label{sec:proper_motion}

We utilized astrometric data from {\it Gaia} Early Data Release 3  \citep[EDR3;][]{Gaia_Brown_2021A&A...649A...1G} to compute the systemic proper motion of Eri~IV. 
We first filtered the sample by removing stars with non-zero parallax ($\varpi-3 \sigma_\varpi>0$) and large proper motion ($v_{\rm tan} - 3 \sigma_{v_{\rm tan}} > v_{\rm esc}$). We applied the following {\it Gaia} quality flags: \texttt{ruwe}$<1.4$, \texttt{astrometric\_excess\_noise\_{\allowbreak}sig} $<2$, $|C^*| \le 3 \sigma_{C^*}(G)$, \texttt{duplicated\_{\allowbreak}source}=False, \texttt{astrometric\_params\_{\allowbreak}solved} $>3$ \citep{Gaia_Lindegren_2021A&A...649A...2L, Gaia_Riello2021A&A...649A...3R}.
We removed background AGN/QSOs that are cross-matched to the {\it Gaia} catalog using \texttt{gaiaedr3.agn\_cross\_id}.
We constructed a Gaussian mixture model composed of a satellite and Milky Way background component fit to the proper motions and spatial positions of stars after applying a color--magnitude filter \citep{Pace2019ApJ...875...77P}.  
We generally followed the methodology of \citet{Pace2019ApJ...875...77P}, but instead of fixing the spatial parameters of Eri~IV, we assumed Gaussian priors based on the best-fit parameters from the \code{ugali} analysis as was done in \citet{delve2}. 

We derived a systemic proper motion of $\mu_{\alpha} \cos \delta= +0.25\pm 0.06 \mas \yr^{-1}$  and  $\mu_{\delta} =-0.10\pm 0.05 \mas \yr^{-1}$, and we found that the sum of the mixture model membership probabilities is $\sum_i p_{i, {\rm MM}} =31.5\pm3.7$. 
In the right panel of \figref{membership}, we present a proper motion scatterplot of all stars meeting the cuts described above in gray, and overlay the likely members identified by the mixture model in colors representing mixture-model-derived probabilities.  In the center panel, we circle these stars in the DELVE color--magnitude diagram. The clear overdensity of stars with consistent proper motions further indicates that Eri~IV is a co-moving system.

\section{Discussion}
\label{sec:discussion}
\subsection{Classification of Eridanus IV}
In \figref{population}, we show the best-fit parameters of Eri~IV in the plane of absolute $V$-band magnitude and physical half-light radius ($M_V - r_{1/2}$) along with other known Milky Way satellites, including classical globular clusters, faint halo star clusters, and candidate and confirmed dwarf galaxies \citep[see][and references therein]{delve2}. The position of Eri~IV in this space is consistent with the known population of ultra-faint dwarf galaxies, suggesting a dwarf galaxy classification for the system. In particular, the large physical size ($r_{1/2} = 75 \pc$) and ellipticity ($\ellip=0.54$) of Eri~IV are inconsistent with the population of comparably bright globular clusters \citep[][2010 edition]{Harris:1996}.Spectroscopic measurement of a large velocity dispersion for this system could provide evidence that the system is a dark-matter-dominated dwarf galaxy. Furthermore, spectroscopic measurements of a metallicity dispersion could provide independent evidence for the system's classification, since a large dispersion could imply that Eri IV has undergone multiple generations of star formation, which is more suggestive of a dwarf galaxy nature \citep[e.g,][]{2012AJ....144...76W}.  

\par We also note that the predicted orbital properties of Eri~IV based on sky position, distance, proper motion, and a range of possible line-of-sight velocities are slightly more consistent with the known population of dwarf galaxies than globular clusters (see \secref{orbit_section}), although the system's morphological properties provide stronger evidence for its classification.

\subsection{RR Lyrae Variable Stars in Eridanus IV}
The color--magnitude diagram in \figref{membership} shows that the horizontal branch of Eri~IV is well-populated, with several stars residing in the instability strip, where Population II variable stars are likely to be found. In particular, based on the empirical relation from \citet{MartinezVazquez2019}, a dwarf galaxy with $M_V = -4.7$ mag is expected to contain 3 to 4 RR Lyrae (RRL) stars.
\par  We searched the \Gaia DR2 variability catalogs  \citep{Clementini2019A&A...622A..60C, Holl2018A&A...618A..30H} and found two RRL candidates within $20\arcmin$ ($\roughly 6\,r_{\rm h}$). 
However, the mean \Gaia $G$-band magnitudes of these stars are $\gtrsim$ 0.5 mag brighter than the horizontal branch of Eri~IV measured in \Gaia.
The \Gaia variability analysis reports a mean magnitude uncertainty of $\roughly 0.002$ mag with semi-amplitudes of $\sim 0.33$ mag and $\sim 0.34$ mag for these RRL candidates using 33 and 29 ``clean" epochs of observation. The inconsistency between the mean magnitudes of these stars and the horizontal branch of Eri~IV makes it unlikely that these stars are members.
We examined the DELVE photometry for each of these RRL stars and confirmed that they are brighter than the horizontal branch of Eri~IV by at least 0.3 to 0.5 mag. While the mean magnitudes of these RRL stars are inconsistent with the horizontal branch of Eri~IV, the clustering of two distant halo RRL within such a small angular area is also unexpected. It is possible that these stars are incorrectly classified as RRLs.
Another possibility is that these stars are being brightened by unresolved binary companions. More complete variability information from \Gaia DR3 may provide additional insight into the nature of these stars.
\par We also searched the PS1 RRL catalog \citep{2017AJ....153..204S} and found one additional candidate RRL star less than $2\arcmin$ from Eri~IV's centroid, but with an estimated distance modulus of 19.14 mag ($\roughly 0.3$ mag closer than Eri~IV). Inspection of the DELVE photometry for this star reveals that it is consistent with the isochrone locus shown in \figref{membership}, and was assigned a \ugali membership probability of 0.98 and a mixture model membership probability of 0.98\footnote{The Gaia EDR3 $\var{source\_id}$ for this star is 3182724869762754048.}, although we note that the former depends on the mean magnitude of this object, which is the sparsely sampled by DELVE. The \Gaia $G$-band mean magnitude of this object, which is reported in $\Gaia$ EDR3 based on 22 epochs, is also consistent with isochrone locus of Eri~IV. Thus, it seems likely that this RRL candidate is a member of Eri IV, despite the inconsistent distance modulus reported by \citet{2017AJ....153..204S}.

\par Lastly,  we note that both the existing RRL catalogs are incomplete at faint magnitudes \citep{Clementini2019A&A...622A..60C,2017AJ....153..204S}, and thus may not include all RRL members of this system. Given the sparse sampling of the DELVE data at the location of Eri~IV, even when augmented by the existing PS1 data, we do not attempt to use it to identify RRL and instead note that follow-up observations are merited to identify and characterize variable stars in this system. 

\begin{figure*}
    \centering
    \includegraphics[width = .99\textwidth]{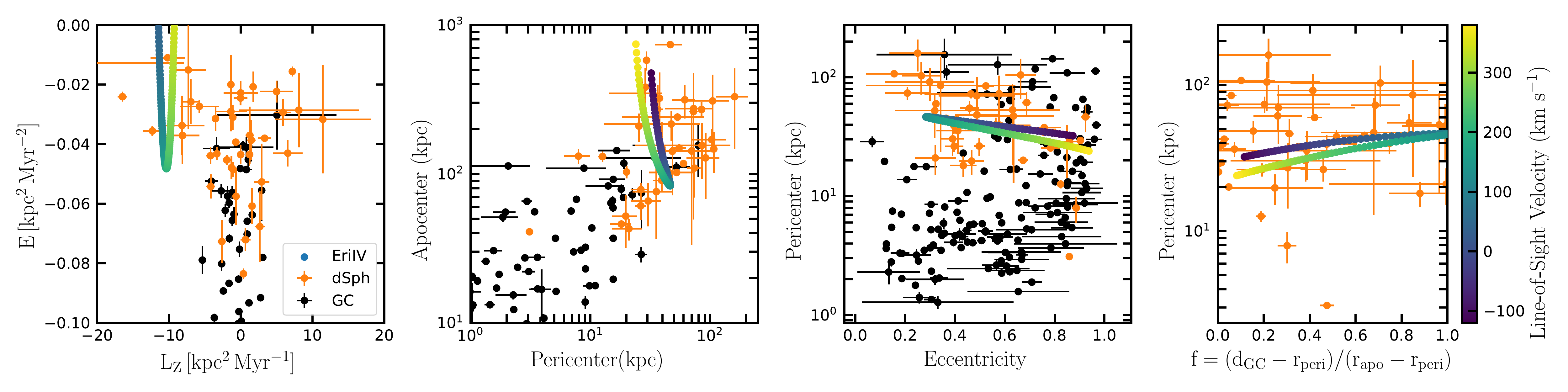}
    \caption{(Left) Orbital energy vs. angular momentum of Eri~IV in the Galactic $z$-direction compared to other dwarf spheroidal galaxies (orange) and globular clusters (black).  The color bar shows the predictions for different line-of-sight velocity values. (Center Left) Apocenter vs.\ pericenter. (Center Right) Pericenter vs.\ eccentricity. (Right) Pericenter vs.\ orbital phase.}
    \label{fig:orbit}
\end{figure*}

\subsection{Orbit}
\label{sec:orbit_section}
While the  line-of-sight velocity of Eri~IV is currently unknown, we can explore possible orbits  by scanning over a range of line-of-sight velocities.
We used the \texttt{gala} package to compute orbits with the \texttt{MilkyWayPotential} Galactic potential \citep{gala}, which consists of two \citet{Hernquist1990ApJ...356..359H} spheroids to model the bulge and nucleus, a \citet{Miyamoto1975PASJ...27..533M} axisymmetric stellar disk, and a NFW dark matter halo \citep{Navarro1996ApJ...462..563N}.  
For a given heliocentric line-of-sight velocity, $v_{\rm los}$, we computed the integrals of motion, $E$ and $L_z$, and the orbital  pericenter ($r_{\rm peri}$), apocenter ($r_{\rm apo}$), and eccentricity.
For comparison, we computed the same properties for other Milky Way satellite galaxies and globular clusters using the 6D phase space information compiled in Pace et al.\ (in prep) and \citet{Vasiliev2021arXiv210209568V}.
In \figref{orbit}, we show the orbital properties of Eri~IV and the population of other known satellites.
For values of $-180 \leq v_{\rm los} \leq 420 \kms$, we find that Eri~IV is bound to the Milky Way. 
\par Interestingly, the observed Milky Way satellite galaxies are preferentially found near their orbital pericenter whereas basic orbital dynamics predicts that satellites on elliptical orbits should spend more time near their apocenters \citep{Fritz2018A&A...619A.103F, Li2021arXiv210403974L}. 
To compare Eri~IV to the known population of satellite galaxies, we examined the ratio, $f = (d_{\rm GC} − r_{\rm peri})/(r_{\rm apo} − r_{\rm peri})$, where $d_{\rm GC}$ is the distance to the Galactic center. 
The ratio $f$ quantifies whether a satellite is close to its pericenter ($f=0$) or apocenter ($f=1$). 
As pointed out by \citet{Fritz2018A&A...619A.103F}, there is an observed excess of satellites with $f < 0.1$; in particular, 16/30 satellites with high-quality proper motion measurements from \Gaia DR2 were found to have $f < 0.1$, which they identify as a possible observational selection effect.
In \figref{orbit}, we find that Eri~IV is only near its pericenter ($f<0.25$) if it has a large eccentricity, which corresponds to the velocity extremes of the range we explored ($v_{\rm los}\lesssim-70 \kms$ or $v_{\rm los} \gtrsim 315 \kms$). 
Thus, Eri~IV may be consistent with the prediction that many satellites will be discovered near their orbital apocenters \citep{Fritz2018A&A...619A.103F}.

\subsection{Potential Tidal Feature} 
\label{sec:tidal}
As seen in the leftmost panel of \figref{diagnostic}, a clear overdensity of isochrone-selected stars is visible to the northeast of the densest region of Eri IV.  This overdensity is also independently visible (albeit less prominent) in members selected solely on the basis of color and \Gaia proper motion, supporting its characterization as a real structure. We note that despite the presence of Galactic cirrus, the $E(B-V)$ measured by \citet{Schlegel:1998} varies by $\lesssim 0.02$ over the Eri~IV field, making it unlikely that this feature is artificially introduced by spatial variations in the foreground dust. 
\par Such extended stellar features can arise via tidal stripping of satellite member stars via interactions with the Milky Way \citep[e.g.,][]{2012MNRAS.420.2700K}, but should be cautiously interpreted in systems like Eri IV where the total number of observed stars is low.  While we defer more detailed spatial modeling and analysis of this candidate tidal feature until deeper photometric data are available, we investigated the possibility that our parameter estimates are biased due to the asymmetric spatial distribution produced by this overdensity. A similar situation was considered by \citet{Drlica-Wagner:2015} for the tidally disrupting satellite Tucana III, and we followed their procedure by fitting an azimuthally symmetric ($\ellip = 0$) stellar density profile. 
We find that the azimuthally symmetric fit of Eri~IV yields a consistent value of the azimuthally-averaged half-light radius ($r_{\rm h}$) within the quoted uncertainty. 
This suggests that the potential tidal feature is not significantly inflating the best-fit half-light radius of Eri~IV (and thereby biasing the system's classification); however, we do observe that the best-fit centroid of Eri~IV is slightly offset from the location of highest density (\figref{diagnostic}). 
\par We note that the proper motion vector of the system is aligned nearly perpendicular to the potential tidal feature and the major axis of Eri~IV (\figref{diagnostic}). Whether or not this configuration favors or disfavors tidal disruption as the cause of the feature depends on the orientation of Eri~IV's orbital plane and phase \citep[e.g.,][]{2012MNRAS.420.2700K}.
\par Lastly, we note that our orbit model predicts that Eri~IV has a pericenter of $r_{\rm peri}>30\kpc$, which is more distant than many other tidally disrupting systems (e.g., Palomar 5, Sagittarius, Tucana III). 
While hints of tidal disruption have been suggested for satellites with larger pericenter distances (e.g., Hercules, Leo V, and Canes Venatici II; \citealt{2012ApJ...756...79S,2015ApJ...804..134R}), some of these claims have been called into question by deeper $\textit{Hubble Space Telescope}$ observations \citep{2019ApJ...885...53M,2020ApJ...902..106M}.
Furthermore, \citet{Ji:2021} recently emphasized the important gravitational influence of the LMC when modeling the orbit of the tidally disturbed system Antlia II.
If Eri IV is indeed experiencing tidal disruption, it could provide valuable insight into the process of satellite disruption in the outer Milky Way halo. Deeper photometry and radial velocity measurements are critical to confirm this candidate tidal feature and to determine the full orbital characteristics of Eri IV.

\subsection{Association with Local Group structures}
\par While a number of recently discovered ultra-faint dwarf galaxies are likely to be presently associated with the Magellanic Clouds \citep[e.g.,][]{Erkal2020MNRAS.495.2554E, Patel2020ApJ...893..121P}, this is unlikely for Eri~IV based on its current location. Eri~IV is located $\roughly 68$ kpc and $\roughly 83$ kpc from the LMC and SMC, respectively, and is not aligned with the Magellanic Stream. Nonetheless, our orbit model does not rule out the possibility of a past association between Eri~IV and the LMC, and a more conclusive investigation will require radial velocity information.
\par Eri~IV is situated close in projection to the Sagittarius (Sgr) stream, leading us to investigate a possible association between the two systems. 
Inspecting the Sgr stream model from \citet{2021MNRAS.501.2279V}, we find that a wrap of the Sgr stream passes to the north of Eri~IV at a heliocentric distance of $\roughly 50$ kpc. However, this is not especially well-matched to the distance of Eri~IV.
Furthermore, we note that the measured proper motion of Eri~IV is misaligned with the modeled proper motion of the Sgr stream at this location.
Thus, we find no obvious connection between Eri~IV and the Sgr stream. Likewise, we find that Eri IV is situated close in projection to the Gj{\"o}ll stellar stream \citep[e.g.,][]{2020ApJ...901...23H}, but note that Eri IV is $>$ 40 kpc more distant than Gj{\"o}ll's apocenter and has an inconsistent proper motion.

\par Lastly, it has been proposed that a substantial fraction of the observed  satellite galaxies co-orbit the Milky Way in a thin plane known as the ``Vast Polar Structure'' (\citealt{2012MNRAS.423.1109P, Fritz2018A&A...619A.103F}). 
We find that the spatial position of Eri~IV places it outside of the plane of this postulated structure.

\section{Summary} 
\label{sec:conclusion}
We have presented the discovery of Eridanus~IV, a new ultra-faint dwarf galaxy candidate, and we have characterized its morphology, stellar population, and proper motion using DELVE photometry and \Gaia astrometry. We have briefly investigated spatially coincident RR Lyrae variable stars, predicted orbital properties, a possible signature of tidal disruption, and potential associations with other Local Group structures. Further characterization of Eri~IV will benefit from additional photometric and spectroscopic data.
\par The discovery of yet another ultra-faint satellite in a region of the sky that was previously observed by shallower surveys is consistent with the prediction that hundreds of ultra-faint galaxies remain to be discovered with surveys like DELVE, HSC-SSP, and the Rubin Observatory LSST.  The discovery and study of these satellites will play an important role in our understanding of the first galaxies, the formation of the Milky Way halo, and the nature of dark matter.

\section{Acknowledgments}
The DELVE project is partially supported by Fermilab LDRD project L2019-011 and the NASA Fermi Guest Investigator Program Cycle 9 No. 91201.
ABP acknowledges support from NSF grant AST-1813881.
This research received support from the National Science Foundation (NSF) under grant no.\ NSF DGE-1656518 through the NSF Graduate Research Fellowship received by SM.
BMP is supported by an NSF Astronomy and Astrophysics Postdoctoral Fellowship under award AST-2001663.
JLC acknowledges support from NSF grant AST-1816196.
JDS acknowledges support from NSF grant AST-1714873.
SRM acknowledges support from NSF grant AST-1909497.
DMD acknowledges
financial support from the Talentia Senior Program (through the
incentive ASE-136) from Secretar\'\i a General de  Universidades,
Investigaci\'{o}n y Tecnolog\'\i a, de la Junta de Andaluc\'\i a.

This project used data obtained with the Dark Energy Camera (DECam), which was constructed by the Dark Energy Survey (DES) collaboration.
Funding for the DES Projects has been provided by 
the DOE and NSF (USA),   
MISE (Spain),   
STFC (UK), 
HEFCE (UK), 
NCSA (UIUC), 
KICP (U. Chicago), 
CCAPP (Ohio State), 
MIFPA (Texas A\&M University),  
CNPQ, 
FAPERJ, 
FINEP (Brazil), 
MINECO (Spain), 
DFG (Germany), 
and the collaborating institutions in the Dark Energy Survey, which are
Argonne Lab, 
UC Santa Cruz, 
University of Cambridge, 
CIEMAT-Madrid, 
University of Chicago, 
University College London, 
DES-Brazil Consortium, 
University of Edinburgh, 
ETH Z{\"u}rich, 
Fermilab, 
University of Illinois, 
ICE (IEEC-CSIC), 
IFAE Barcelona, 
Lawrence Berkeley Lab, 
LMU M{\"u}nchen, and the associated Excellence Cluster Universe, 
University of Michigan, 
NSF's National Optical-Infrared Astronomy Research Laboratory, 
University of Nottingham, 
Ohio State University, 
OzDES Membership Consortium
University of Pennsylvania, 
University of Portsmouth, 
SLAC National Lab, 
Stanford University, 
University of Sussex, 
and Texas A\&M University.

This work has made use of data from the European Space Agency (ESA) mission {\it Gaia} (\url{https://www.cosmos.esa.int/gaia}), processed by the {\it Gaia} Data Processing and Analysis Consortium (DPAC, \url{https://www.cosmos.esa.int/web/gaia/dpac/consortium}).
Funding for the DPAC has been provided by national institutions, in particular the institutions participating in the {\it Gaia} Multilateral Agreement.

Based on observations at Cerro Tololo Inter-American Observatory, NSF's National Optical-Infrared Astronomy Research Laboratory (2019A-0305; PI: Drlica-Wagner), which is operated by the Association of Universities for Research in Astronomy (AURA) under a cooperative agreement with the National Science Foundation.

This manuscript has been authored by Fermi Research Alliance, LLC, under contract No.\ DE-AC02-07CH11359 with the US Department of Energy, Office of Science, Office of High Energy Physics. The United States Government retains and the publisher, by accepting the article for publication, acknowledges that the United States Government retains a non-exclusive, paid-up, irrevocable, worldwide license to publish or reproduce the published form of this manuscript, or allow others to do so, for United States Government purposes.

\facility{Blanco, \Gaia.}
\software{\emcee \citep{Foreman-Mackey:2013}, \code{gala} \citep{gala}, \healpix \citep{Gorski:2005},\footnote{\url{http://healpix.sourceforge.net}} \code{healpy},\footnote{\url{https://github.com/healpy/healpy}} \ugali \citep{Bechtol:2015}.\footnote{\url{https://github.com/DarkEnergySurvey/ugali}}}

\bibliography{main}

\appendix
\section{Posterior Distributions of Eridanus IV Parameters}
\label{app:posterior}
\begin{figure}[H]
    \centering
    \includegraphics[width = .95\textwidth]{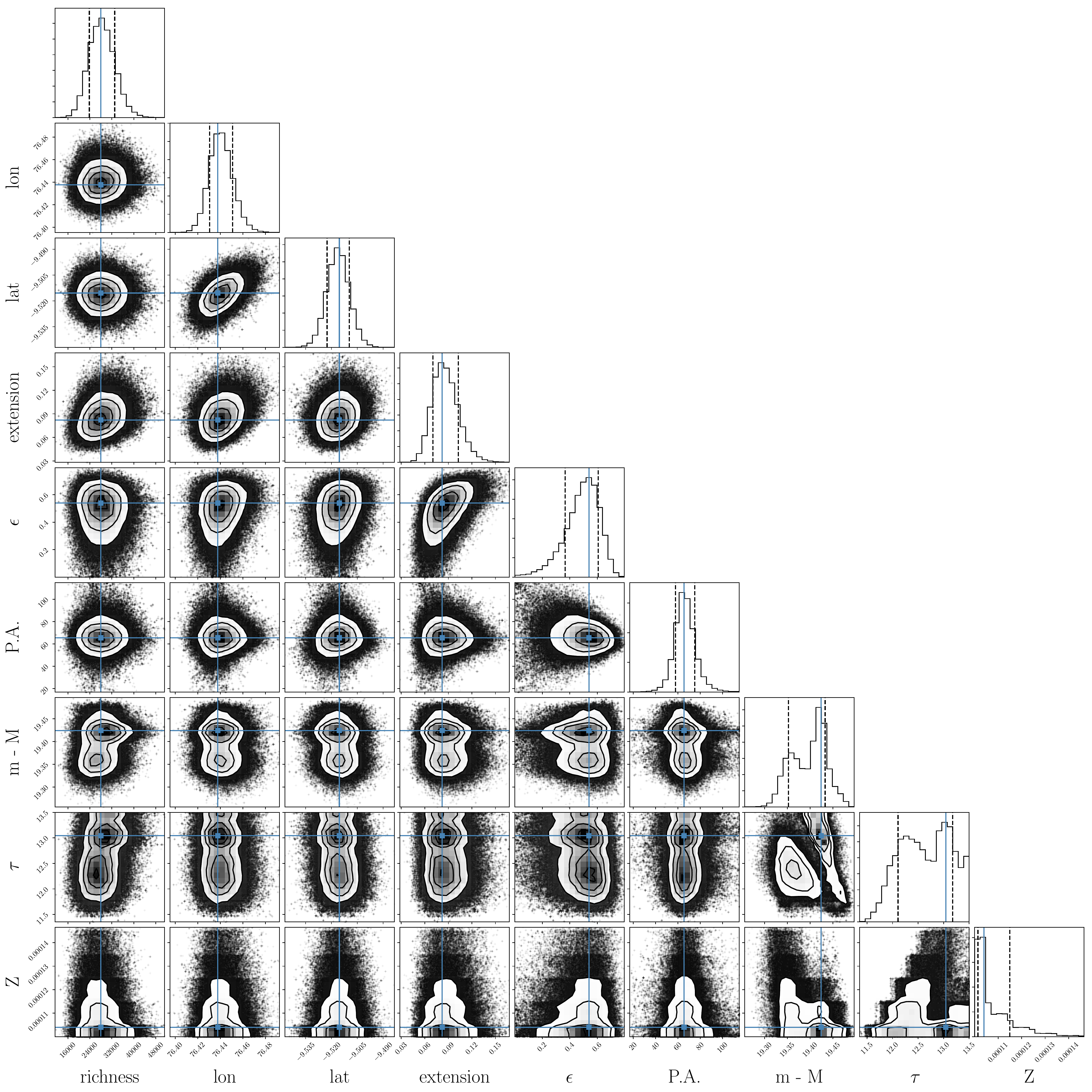}
    \caption{Posterior probability distributions for each parameter, derived from the MCMC sampling described in \secref{results}. The distance modulus and age parameters display bimodality, leading to asymmetric errors associated with these parameters in particular. }
    \label{fig:posterior}
\end{figure}

\end{document}

%% file: commands.tex
\usepackage{amsmath}
\usepackage{amssymb}
\usepackage{xspace}
\usepackage{xifthen}
\usepackage{eso-pic}

\newcommand{\Gaia}{{\it Gaia}\xspace}

\definecolor{forestgreen}{HTML}{228B22}
\definecolor{urlblue}{HTML}{000000}

\mathchardef\mhyphen="2D

\newcommand{\roughly}{\ensuremath{ {\sim}\,} }

\newlength{\dhatheight}

\newcommand{\code}[1]{\texttt{#1}\xspace}

\newcommand{\unit}[1]{\ensuremath{\mathrm{\,#1}}\xspace}
\newcommand{\yr}{\unit{yr}}
\newcommand{\Gyr}{\unit{Gyr}}

\newcommand{\degree}{\ensuremath{{}^{\circ}}\xspace}

\newcommand{\mas}{\unit{mas}}

\newcommand{\km}{\unit{km}}
\newcommand{\kms}{\km \second^{-1}}
\newcommand{\pc}{\unit{pc}}
\newcommand{\kpc}{\unit{kpc}}
\newcommand{\Mpc}{\unit{Mpc}}
\newcommand{\second}{\unit{s}}

\newcommand{\Msun}{\unit{M_\odot}}

\newcommand{\magn}{\unit{mag}}

\newcommand{\secref}[1]{Section~\ref{sec:#1}}
\newcommand{\appref}[1]{Appendix~\ref{app:#1}}
\newcommand{\tabref}[1]{Table~\ref{tab:#1}}

\newcommand{\figref}[1]{Figure~\ref{fig:#1}}

\newcommand{\bandvar}[2][]{%
  \ifthenelse{\isempty{#1}}{\var{#2}}{\var{#2\_#1}}%
}

\newcommand{\LCDM}{\ensuremath{\rm \Lambda CDM}\xspace}
\newcommand{\modulus}{\ensuremath{m - M}\xspace}

\newcommand{\ra}{{\ensuremath{\alpha_{2000}}}\xspace}
\newcommand{\dec}{{\ensuremath{\delta_{2000}}}\xspace}
\newcommand{\age}{{\ensuremath{\tau}}\xspace}
\newcommand{\metal}{{\ensuremath{Z}}\xspace}

\newcommand{\feh}{{\ensuremath{\rm [Fe/H]}}\xspace}
\newcommand{\ellip}{\ensuremath{\epsilon}\xspace}
\newcommand{\PA}{\ensuremath{\mathrm{P.A.}}\xspace}
\newcommand{\TS}{\ensuremath{\mathrm{TS}}\xspace}

\newcommand{\HEALPix}{\code{HEALPix}}
\newcommand{\healpix}{\HEALPix}

\newcommand{\emcee}{\code{emcee}}
\newcommand{\ugali}{\code{ugali}}

\newcommand{\var}[1]{\ensuremath{\texttt{\MakeUppercase{#1}}}\xspace}

\providecommand\physrep{\ref@jnl{Phys.~Rep.}}%
\providecommand\apjs{\ref@jnl{ApJS}}%
\providecommand{\jcap}{\ref@jnl{JCAP}}%

%% file: authors.tex

\author[0000-0003-1697-7062]{W.~Cerny}
\affiliation{Kavli Institute for Cosmological Physics, University of Chicago, Chicago, IL 60637, USA}
\affiliation{Department of Astronomy and Astrophysics, University of Chicago, Chicago IL 60637, USA}

\author[0000-0002-6021-8760]{A.~B.~Pace}
\affiliation{McWilliams Center for Cosmology, Carnegie Mellon University, 5000 Forbes Avenue, Pittsburgh, PA 15213, USA}

\author[0000-0001-8251-933X]{A.~Drlica-Wagner}
\affiliation{Fermi National Accelerator Laboratory, P.O.\ Box 500, Batavia, IL 60510, USA}
\affiliation{Kavli Institute for Cosmological Physics, University of Chicago, Chicago, IL 60637, USA}
\affiliation{Department of Astronomy and Astrophysics, University of Chicago, Chicago IL 60637, USA}

\author[0000-0003-2644-135X]{S.~E.~Koposov}
\affiliation{Institute for Astronomy, University of Edinburgh, Royal Observatory, Blackford Hill, Edinburgh EH9 3HJ, UK}
\affiliation{Institute of Astronomy, University of Cambridge, Madingley Road, Cambridge CB3 0HA, UK}
\affiliation{Kavli Institute for Cosmology, University of Cambridge, Madingley Road, Cambridge CB3 0HA, UK}

\author[0000-0003-4341-6172]{A.~K.~Vivas}
\affiliation{Cerro Tololo Inter-American Observatory, NSF's National Optical-Infrared Astronomy Research Laboratory,\\ Casilla 603, La Serena, Chile}

\author[0000-0003-3519-4004]{S.~Mau}
\affiliation{Department of Physics, Stanford University, 382 Via Pueblo Mall, Stanford, CA 94305, USA}
\affiliation{Kavli Institute for Particle Astrophysics \& Cosmology, P.O.\ Box 2450, Stanford University, Stanford, CA 94305, USA}

\author[0000-0002-7134-8296]{A.~H.~Riley}
\affiliation{George P. and Cynthia Woods Mitchell Institute for Fundamental Physics and Astronomy, Texas A\&M University, College Station, TX 77843, USA}
\affiliation{Department of Physics and Astronomy, Texas A\&M University, College Station, TX 77843, USA}

\author[0000-0003-4383-2969]{C.~R.~Bom}
\affiliation{Centro Brasileiro de Pesquisas F\'isicas, Rua Dr. Xavier Sigaud 150, 22290-180 Rio de Janeiro, RJ, Brazil}

\author[0000-0002-3936-9628]{J.~L.~Carlin}
\affiliation{Rubin Observatory/AURA, 950 North Cherry Avenue, Tucson, AZ, 85719, USA}

\author[0000-0003-1680-1884]{Y.~Choi}
\affiliation{Space Telescope Science Institute, 3700 San Martin Drive, Baltimore, MD 21218, USA}

\author[0000-0002-8448-5505]{D.~Erkal}
\affiliation{Department of Physics, University of Surrey, Guildford GU2 7XH, UK}

\author[0000-0001-6957-1627]{P.~S.~Ferguson}
\affiliation{George P. and Cynthia Woods Mitchell Institute for Fundamental Physics and Astronomy, Texas A\&M University, College Station, TX 77843, USA}
\affiliation{Department of Physics and Astronomy, Texas A\&M University, College Station, TX 77843, USA}

\author[0000-0001-5160-4486]{D.~J.~James}
\affiliation{Center for Astrophysics, Harvard \& Smithsonian, 60 Garden Street, Cambridge, MA 02138, USA}
\affiliation{ASTRAVEO, LLC, PO Box 1668, Gloucester, MA 01931}

\author[0000-0002-9110-6163]{T.~S.~Li}
\affiliation{Observatories of the Carnegie Institution for Science, 813 Santa Barbara Street, Pasadena, CA 91101, USA}
\affiliation{Department of Astrophysical Sciences, Princeton University, Princeton, NJ 08544, USA}

\author{D.~Mart\'{i}nez-Delgado}
\affiliation{Instituto de Astrof\'{i}sica de Andaluc\'{i}a, CSIC, E-18080 Granada, Spain}

\author[0000-0002-9144-7726]{C.~E.~Mart\'inez-V\'azquez}
\affiliation{Cerro Tololo Inter-American Observatory, NSF's National Optical-Infrared Astronomy Research Laboratory, Casilla 603, La Serena, Chile}

\author{R.~R.~Munoz}
\affiliation{Departamento de Astronom\'ia, Universidad de Chile, Camino El Observatorio 1515, Las Condes, Santiago, Chile}

\author[0000-0001-9649-4815]{B.~Mutlu-Pakdil}
\affiliation{Kavli Institute for Cosmological Physics, University of Chicago, Chicago, IL 60637, USA}
\affiliation{Department of Astronomy and Astrophysics, University of Chicago, Chicago IL 60637, USA}

\author[0000-0002-7134-8296]{K.~A.~G.~Olsen}
\affiliation{NSF's National Optical-Infrared Astronomy Research Laboratory, 950 N. Cherry Ave., Tucson, AZ 85719, USA}

\author[0000-0001-9186-6042]{A.~Pieres}
\affiliation{Laborat\'orio Interinstitucional de e-Astronomia - LIneA, Rua Gal. Jos\'e Cristino 77, Rio de Janeiro, RJ - 20921-400, Brazil }
\affiliation{Observat\'orio Nacional, Rua Gal. Jos\'e Cristino 77, Rio de Janeiro, RJ - 20921-400, Brazil}

\author[0000-0002-1594-1466]{J.~D.~Sakowska}
\affiliation{Department of Physics, University of Surrey, Guildford GU2 7XH, UK}

\author[0000-0003-4102-380X]{D.~J.~Sand}
\affiliation{Department of Astronomy/Steward Observatory, 933 North Cherry Avenue, Room N204, Tucson, AZ 85721-0065, USA}

\author[0000-0002-4733-4994]{J.~D.~Simon}
\affiliation{Observatories of the Carnegie Institution for Science, 813 Santa Barbara Street, Pasadena, CA 91101, USA}

\author[0000-0003-2599-7524]{A.~Smercina}
\affiliation{Astronomy Department, University of Washington, Box 351580, Seattle, WA 98195-1580, USA}

\author[0000-0003-1479-3059]{G.~S.~Stringfellow}
\affiliation{Center for Astrophysics and Space Astronomy, University of Colorado, 389 UCB, Boulder, CO 80309-0389, USA}

\author[0000-0002-9599-310X]{E.~J.~Tollerud}
\affiliation{Space Telescope Science Institute, 3700 San Martin Drive, Baltimore, MD 21218, USA}

\author[0000-0002-6904-359X]{M.~Adam\'ow}
\affiliation{National Center for Supercomputing Applications, University of Illinois, 1205 West Clark Street, Urbana, IL 61801, USA}

\author{D.~Hernandez-Lang}
\affiliation{Ludwig-Maximilians-Universität München, Scheinerstraße 1, München, Germany}

\author[0000-0003-2511-0946]{N.~Kuropatkin}
\affiliation{Fermi National Accelerator Laboratory, P.O.\ Box 500, Batavia, IL 60510, USA}

\author[0000-0003-3402-6164]{L.~Santana-Silva}
\affiliation{NAT-Universidade Cruzeiro do Sul / Universidade Cidade de S{\~a}o Paulo, Rua Galv{\~a}o Bueno, 868, 01506-000, S{\~a}o Paulo, SP, Brazil}

\author[0000-0001-7211-5729]{D.~L.~Tucker}
\affiliation{Fermi National Accelerator Laboratory, P.O.\ Box 500, Batavia, IL 60510, USA}

\author[0000-0001-6455-9135]{A.~Zenteno}
\affiliation{Cerro Tololo Inter-American Observatory, NSF's National Optical-Infrared Astronomy Research Laboratory,\\ Casilla 603, La Serena, Chile}

\correspondingauthor{William Cerny}
\email{williamcerny@uchicago.edu}

\collaboration{(DELVE Collaboration)}

%% file: Table1.tex
\begin{deluxetable}{l c c}[H]
\tablecolumns{3}
\tablewidth{0pt}
\tabletypesize{\footnotesize}
\tablecaption{\label{tab:properties} Morphological, isochrone, and proper motion parameters for \stamp.}
\tablehead{
\colhead{Parameter} & \colhead{Value}   & \colhead{Units}}
\startdata
\ra   & $76.438^{+0.012}_{-0.008}$ & deg \\
\dec  & $-9.515^{+0.006}_{-0.007}$ & deg\\
$a_\text{h}$ & $4.9^{+1.1}_{-0.8}$ &  arcmin \\
$r_\text{h}$  & $3.3^{+0.7}_{-0.6}$ &  arcmin  \\
$r_{1/2}$  & $75^{+16}_{-13}$ & pc  \\
\ellip & $0.54^{+0.10}_{-0.14}$ & ... \\
\PA  & $65^{+9}_{-8}$ & deg \\
\modulus  &  $19.42^{+ 0.01}_{-0.08} \pm 0.1$\tablenotemark{a} & mag\\
$D_{\odot}$  & $76.7^{+4.0}_{-6.1}$  & kpc\\
\age  & $13.0^{+0.1}_{-1.0}$ & Gyr \\
\metal & $<0.00013$\tablenotemark{b} & ... \\
$\sum_i p_{i, {\rm \ugali}}$  & $69^{+9}_{-9}$ & ... \\
\TS & 200.6 & ... \\[-0.5em]
\multicolumn{3}{c}{\hrulefill} \\
$M_V$ & $-4.7 \pm 0.2$\tablenotemark{c} & mag  \\
$\mu$  & 28.2 & mag~arcsec$^{-2}$\\
$M_{*}$ & $6519^{+1038}_{-1111}$ 
& ${\rm M}_{\odot}$ \\
\feh  &$< -2.08$\tablenotemark{b}  &  dex \\
$E(B-V)$ & 0.107\tablenotemark{d} & mag \\
$\ell$ & 209.499 &deg \\
$b$  &  -27.772 & deg \\
\multicolumn{3}{c}{\hrulefill} \\
$\mu_{\alpha} \cos \delta$  & $+0.25\pm 0.06$ &  mas~yr$^{-1}$  \\
$\mu_{\delta}$ & $-0.10\pm 0.05$ &  mas~yr$^{-1}$ \\
$\sum_i p_{i, {\rm MM}}$ & $31.5\pm3.7$ &  ...\\[+0.5em]
\enddata
\tablecomments{The quoted uncertainties were derived from the highest density interval containing the peak and 68\% of the marginalized posterior distribution.}
\tablenotetext{a}{We assume a systematic uncertainty of $\pm0.1$ associated with isochrone modeling \citep{Drlica-Wagner:2015}.}

\tablenotetext{b}{The metallicity posterior peaked at the lower bound of the allowed parameter range ($Z$ = 0.0001);  we  therefore quote an upper limit at the 95\% confidence level.}

\tablenotetext{c}{The uncertainty in $M_V$ was calculated following \citet{Martin:2008} and does not include uncertainty in the distance.}

\tablenotetext{d}{This $E(B-V)$ value refers to the mean reddening of all sources within one half-light radius.}

\vspace{-3em}
\end{deluxetable}